\documentclass[conference]{IEEEtran}

\usepackage{soul}
\usepackage{float}
\usepackage{url}
\usepackage{tabularx}
\usepackage{array} 
\usepackage{makecell}
\usepackage{multirow}
\usepackage{booktabs}
\usepackage[colorlinks,urlcolor=blue,linkcolor=blue,citecolor=blue]{hyperref}
\usepackage[dvipsnames,table]{xcolor}
\usepackage{xspace}
\usepackage{rotating}
\usepackage{tabularx}
\definecolor{LightCyan}{rgb}{0.88,1,1}
\usepackage{balance}
\usepackage{graphicx} 
\usepackage{adjustbox} 
\usepackage{comment}
\usepackage{cite}
\usepackage{amsmath,amssymb,amsfonts}
\usepackage{algorithmic}
\usepackage{cleveref}
\usepackage{listings}

\definecolor{codegreen}{rgb}{0,0.6,0}
\definecolor{codegray}{rgb}{0.5,0.5,0.5}
\definecolor{codepurple}{rgb}{0.58,0,0.82}
\definecolor{backcolour}{rgb}{0.95,0.95,0.92}
\definecolor{commentsColor}{rgb}{0.497495, 0.497587, 0.497464}
\definecolor{keywordsColor}{rgb}{0.000000, 0.000000, 0.635294}
\definecolor{stringColor}{rgb}{0.558215, 0.000000, 0.135316}

\lstdefinestyle{mystyle}{
  backgroundcolor=\color{backcolour},   
  basicstyle=\footnotesize,        
  breakatwhitespace=false,         
  breaklines=true,                 
  captionpos=b,                    
  commentstyle=\color{commentsColor}\textit,    
  deletekeywords={...},            
  escapeinside={\%*}{*)},          
  extendedchars=true,              
  frame=tb,                        
  keepspaces=true,                 
  keywordstyle=\color{keywordsColor}\bfseries,       
  language=Python,                 
  otherkeywords={*,...},           
  numbers=left,                    
  numbersep=5pt,                   
  numberstyle=\tiny\color{commentsColor}, 
  rulecolor=\color{black},         
  showspaces=false,                
  showstringspaces=false,          
  showtabs=false,                  
  stepnumber=1,                    
  stringstyle=\color{stringColor}, 
  tabsize=2,                     
  title=\lstname,                  
  columns=fixed                    
}
\lstdefinelanguage{YARA}{
  keywords={rule, meta, strings, condition, matches, rules, externals},
  keywordstyle=\color{blue}\bfseries,
  ndkeywords={and, match, callback},
  ndkeywordstyle=\color{darkgray}\bfseries,
  identifierstyle=\color{black},
  sensitive=false,
  comment=[l]{//},
  morecomment=[s]{/*}{*/},
  commentstyle=\color{purple}\ttfamily,
  stringstyle=\color{red}\ttfamily,
  morestring=[b]',
  morestring=[b]"
}
\newcommand{\G}{\mathcal{G}\xspace}
\newcommand{\V}{\mathcal{V}\xspace}
\newcommand{\E}{\mathcal{E}\xspace}
\newcommand{\Q}{\mathcal{Q}\xspace}
\newcommand{\lb}{\mathcal{L}\xspace}
\newcommand{\pr}{\mathcal{P}\xspace}
\newcommand{\RoE}{\mathcal{R}\xspace}
\newcommand{\R}{\mathcal{R}\xspace}
\newcommand{\RE}{\mathcal{RE}\xspace}
\newcommand{\PT}{\mathcal{T}\xspace}

\newcommand{\p}{\mathcal{C}\xspace}
\newcommand{\M}{\mathcal{M}\xspace}
\newcommand{\A}{\mathcal{A}\xspace}
\newcommand{\U}{\mathcal{U}\xspace}

\DeclareRobustCommand{\IEEEauthorrefmark}[1]{\smash{\textsuperscript{\footnotesize #1}}}

\author{
    \IEEEauthorblockN{
    Damodar Panigrahi\IEEEauthorrefmark{1}\IEEEauthorrefmark{3},
    Raj Patel\IEEEauthorrefmark{2}\IEEEauthorrefmark{4},
    Shaswata Mitra\IEEEauthorrefmark{2}\IEEEauthorrefmark{5},
    Sudip Mittal\IEEEauthorrefmark{2}\IEEEauthorrefmark{6},
    Shahram Rahimi\IEEEauthorrefmark{2}\IEEEauthorrefmark{7}
    }
    \IEEEauthorblockA{\IEEEauthorrefmark{1}Mississippi State University, MS, USA}
    \IEEEauthorblockA{\IEEEauthorrefmark{2}The University of Alabama, AL, USA}
    dp1657@msstate.edu\IEEEauthorrefmark{3}, 
    \{rpatel38\IEEEauthorrefmark{4},
      smitra3\IEEEauthorrefmark{5},
      smittal1\IEEEauthorrefmark{6},
      srahimi1\IEEEauthorrefmark{7}\}@ua.edu
}

\begin{document}

\title{IRSDA: An Agent-Orchestrated Framework for Enterprise Intrusion Response}

\maketitle

\begin{abstract}
Modern enterprise systems face escalating cyber threats that are increasingly dynamic, distributed, and multi-stage in nature. Traditional intrusion detection and response systems often rely on static rules and manual workflows, which limit their ability to respond with the speed and precision required in high-stakes environments. To address these challenges, we present the Intrusion Response System Digital Assistant (IRSDA), an agent-based framework designed to deliver autonomous and policy-compliant cyber defense. IRSDA combines Self-Adaptive Autonomic Computing Systems (SA-ACS) with the Knowledge guided Monitor, Analyze, Plan, and Execute (MAPE-K) loop to support real-time, partition-aware decision-making across enterprise infrastructure.

IRSDA incorporates a knowledge-driven architecture that integrates contextual information with AI-based reasoning to support system-guided intrusion response. The framework leverages retrieval mechanisms and structured representations to inform decision-making while maintaining alignment with operational policies. We assess the system using a representative real-world microservices application, demonstrating its ability to automate containment, enforce compliance, and provide traceable outputs for security analyst interpretation. This work outlines a modular and agent-driven approach to cyber defense that emphasizes explainability, system-state awareness, and operational control in intrusion response.
\end{abstract}


\section{Introduction}

The expansion of internet connectivity and digital systems has reshaped the cybersecurity environment for enterprises worldwide. By 2025, more than 5.5 billion individuals (representing nearly 68\% of the global population) use the internet \cite{statistaFeb2025}. In the United States, 98\% of adults now own a smartphone \cite{mobilefact2025}. This level of connectivity has introduced new efficiencies in communication, service delivery, and digital commerce, but has also increased the risk of cyber intrusions. As organizational systems become more interconnected, they also become more exposed to threats that exploit vulnerabilities across cloud infrastructure, mobile endpoints, and enterprise applications \cite{CISA_list, abcnews_list}. Traditional security systems often separate detection from response. Intrusion Detection Systems (IDS) are commonly used to surface anomalies and alerts, but the subsequent handling of incidents typically depends on static playbooks and human-driven processes. These manual interventions are frequently delayed and are not well suited for managing attacks that evolve rapidly or propagate across complex infrastructures. In many cases, intrusions remain active for extended periods, increasing operational risk and making containment difficult \cite{top11CyberAttacks, cardellini2022intrusion}. While recent IDS approaches have incorporated machine learning for adaptive, real-time detection, response mechanisms have not advanced at the same pace, remaining static and heavily reliant on manual security analysts' real-time orchestration.

To address this gap, we propose the Intrusion Response System Digital Assistant (IRSDA), an integrated framework that connects both intrusion detection and automated intrusion response. IRSDA is designed around the principles of self-adaptive systems and follows the MAPE-K (Monitor, Analyze, Plan, Execute, Knowledge) \cite{lemos2013software} loop to support end-to-end autonomous cyber defense and recovery. It uses agentic AI to coordinate response agents across distributed system partitions, allowing them to retrieve contextual data from a centralized knowledge-graph base and apply actions that are aligned with system policies and operational boundaries.

In order to support agent-based orchestration, IRSDA incorporates a large language model (LLM) that has been fine-tuned on large corpora of cybersecurity texts. This is aimed at supporting contextual reasoning, triage, and explanation of observed system behavior and adopted actions in natural language. To ensure factual relevance and minimize hallucination, graph-based retrieval-augmented generation (RAG) is used to ground model output using real-time enterprise data and dynamic Rules-of-Engagement (ROE) \cite{lewis2020retrieval, rag-2, jang-etal-2024-ignore}. Our main contributions are as follows:
\begin{itemize}
\item We introduce Intrusion Response System Digital Assistant (IRSDA), a modular framework that integrates intrusion detection with automated response via iterative agentic orchestration, a property graph knowledge base, and a cybersecurity-tuned LLM.
\item We adopted a graph-based RAG pipeline that aligns model outputs with enterprise telemetry, system logs, and pre-defined policies.
\item We demonstrate IRSDA through a case study of a microservices system, showcasing an end-to-end workflow from detection to ROE-compliant containment, with an analyst explanation.
\end{itemize}

The remainder of this paper is structured as follows. Section II reviews related work and foundational concepts in intrusion response and self-adaptive systems. Section III describes the architecture and design of IRSDA. Section IV presents evaluation results and discusses implications for agentic, explainable cyber defense. Section V concludes with areas for future research and opportunities for broader deployment.

\section{Background}\label{background}
The evolution of cybersecurity defense has been driven by the increasing complexity and dynamism of modern computing environments. Autonomic Computing Systems (ACS) were proposed as a paradigm for self-managing systems that can adapt to changing operational demands \cite{1160055}. The ACS paradigm is structured around the MAPE-K architecture, which includes distinct phases for Monitoring, Analysis, Planning, Execution, and Knowledge management \cite{lemos2013software}. This architecture enables continuous monitoring, dynamic analysis, and adaptive response to changing conditions. To further enhance adaptability in the face of advanced threats, Self-Adaptive Autonomic Computing Systems (SA-ACS) were introduced. These systems support real-time adjustment of defense strategies and are particularly relevant for security applications \cite{macias2013self, krupitzer2015survey}.

Within cybersecurity, the protection of digital assets depends on effective detection and response to threats such as malware, ransomware, and insider attacks. Intrusion Detection Systems (IDS) provide the first line of defense by analyzing network traffic, system logs, and user behaviors to identify suspicious activity \cite{snort-ips, suricata}. However, IDSs alone often struggle to scale and may generate high false-positive rates, particularly in large-scale deployments \cite{iannucci2020hybrid}. To address these limitations, Intrusion Response Systems (IRS) have been developed to automate and accelerate defensive actions. The IRS determines appropriate responses based on the severity and nature of detected threats \cite{inayat2016intrusion, cardellini2022intrusion}. Recent IRS architectures increasingly employ distributed, agent-based frameworks such as the Autonomous Intelligent Cyber Defense Agents (AICA), which use collaboration and adaptive reasoning to strengthen resilience against sophisticated threats \cite{kott2023autonomous}. Persistent challenges in IRS research include managing large and dynamic state spaces, designing effective response policies when uncertainty exists, and ensuring coordination among agents within distributed environments \cite{nguyen2020deep}.

As cyber attacks continue to grow in sophistication and digital infrastructure expands, there is a clear need for interfaces that can combine automation with user accessibility. Digital assistants, which leverage advances in artificial intelligence (AI), are central to this transition. Large Language Models (LLMs) such as GPT, LLAMA, and T5 have demonstrated significant capabilities in natural language processing and generation \cite{radford2019language, brown2020language, touvron2023llama, raffel2020exploring}. These models can be further specialized through fine-tuning on domain-specific datasets, which allows them to provide tailored support for threat intelligence, incident triage, and compliance documentation in cybersecurity \cite{secPalm2}. Additional techniques, such as prompt engineering and Retrieval-Augmented Generation (RAG), enable LLM-based assistants to incorporate external knowledge and produce context-aware outputs that remain up to date \cite{promptEng}.

By integrating fine-tuned, LLM-based digital assistants into the Intrusion Response System Digital Assistant (IRSDA) framework, security operations can benefit from seamless communication between human operators and adaptive cyber defense agents. This integration improves transparency, provides richer contextual awareness, and enhances the effectiveness of enterprise security practices.

\section{Intrusion Response System Digital Assistant (IRSDA)} \label{irsda-design}
This section describes the internal architecture of the IRSDA framework. We begin by outlining the overall system architecture and the functional roles of each modular component (Section \ref{Sysem-Architecture}). We then examine the integration of enterprise data sources, including system logs, configuration files, security rules, and model training datasets (Section \ref{Knowledg-Graph}). The next subsections discuss the generative AI component, its enterprise-specific customization methods, the design of the retrieval pipeline, and the primary tasks performed by the large language model (Sections \ref{llm}, \ref{rag}, and \ref{llm-tasks}).

\begin{table} [!ht]
    \renewcommand{\arraystretch}{1.20}%
    \caption{IRSKG schema notations.}
    \begin{tabularx}{0.48\textwidth} { 
       >{\centering\arraybackslash}p{0.10\textwidth}|X
       >{\raggedright\arraybackslash}X }
        \hline
            \rowcolor{lightgray} 
            \textbf{Symbol} &  \textbf{Description} \\
        \hline
            $i$ & A \textit{component type} \\
            $\p_i$ & A \textit{partition} corresponding to the $i$-th component type \\
            $i_j$ & The $j$-th \textit{component} of the $i$-th component type\\
            $\p_{i_{j}}$ & The $j$-th \textit{component} of the $i$-th type of the $i$-th partition\\
            $\M$ & The enterprise systems model\\
            $\G$ & The entire Graph database \\
            $\V$ & The set of all vertices in the database. \\ 
            $\E$ & The set of all edges in the database \\ 
            $\V_i$ & The $ith$ vertex. \\ 
            $\E_{i,j}$ & The edge between the $\V_i$ and $\V_j$\\ 
            $\lb(\V_i)$ & The label of $\V_i$. \\ 
            $\pr(\V_i)$ & The property key-value dictionary of $\V_i$\\ 
            $\lb(\E_{i,j})$ & The label of $\E_{i,j}$ edge. \\ 
            $\pr(\E_{i,j})$ & The property key-value dictionary of $\E_{i,j}$ edge\\  
            $\RoE$ & Set of all Rules of Engagement \\  
            $\RoE_i$ & $ith$ Rule of Engagement\\
            $\V_{a|b}(\R_i)$ & Vertex $a$ or $b$ of rule $\RoE_i$\\  
            $\lb(\V_a(\R_i))$ & Label of the Vertex $a$ of $\R_i$\\  
            $\pr(\V_a(\R_i))$ & Property of the Vertex $a$ of $\R_i$\\  
            $\E({\RoE_i})$ & Edge between $\V_a$ and $\V_b$ of $\RoE_i$\\  
            $\lb(\E({\RoE_i}))$ & Label of edge between $\V_a$ and $\V_b$ of $\RoE_i$\\  
            $\pr(\E(\RoE_i))$ & Property of edge between $\V_a$ and $\V_b$ of $\RoE_i$\\  
            $\RoE^{t} \ | \ \RoE^i \in \RoE^{ti}$ & A meta-template that different enterprise systems follow and ultimately all $\RoE_{i}$ are compliant to.\\
            $\RoE^{tk}$ & A template for a specific enterprise system $k$ (e.g. A web enterprise system). \\ 
        $\Q_i$ & A \textit{IRSDA Client posted query} \\
        $\PT_i$ & A \textit{IRSLLM prompt technique template} \\
        $\PT$ & Set of all prompt templates\\
        $\A_i$ & IRSLLM response corresponding to the prompt using $\PT_i$ template\\
        $\U_i$ & IRSLLM summarization response corresponding to the response $\A_i$\\
        $\A$ & The set of responses, $\U_i$ one for each prompt using templates $\PT_i$\\
        \hline  
    \end{tabularx}
    \label{tbl:notation-table}
\end{table}

\subsection{Intrusion Response System Digital Assistant Architecture}  \label{Sysem-Architecture}
\begin{figure*}[ht]
    \centering
    \includegraphics[scale=0.545]{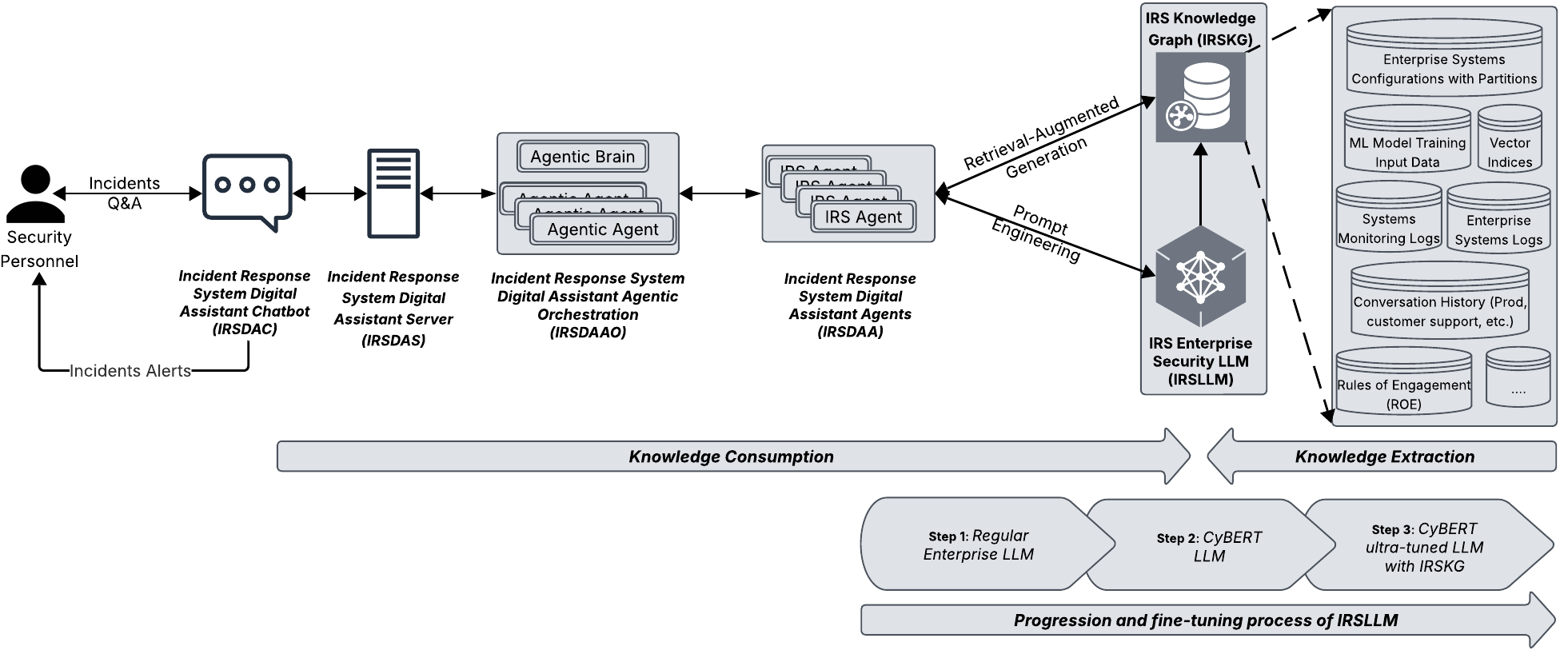}
    \caption{IRSDA System Architecture - an \textit{n-tier} architecture following a client-server and multi-agent system design. Tier I includes the IRSDA Chatbot (\textit{IRSDAC}), the user-facing interface that enables natural language interactions. Tier II is the IRSDA Server (\textit{IRSDAS}), which handles client requests, manages state, and coordinates incident response. Tier III is the IRS Digital Assistant Agentic Orchestration layer (\textit{IRSDAAO}), where an agentic brain orchestrates task delegation to agents. Tier IV contains IRSAgents (\textit{IRSDAA}), with each agent handling a single system partition. Tier IV also includes the IRS Knowledge Graph (\textit{IRSKG}) and the enterprise-tuned LLM (\textit{IRSLLM}). IRSKG is a vector-indexed knowledge base of system logs, configuration data, and rules of engagement. IRSLLM is fine-tuned on both enterprise data and publicly available cybersecurity datasets. The architecture supports both knowledge extraction into IRSKG and consumption by IRSLLM and other system components.}
    \label{fig:irsda-system-architecture}
\end{figure*}

Here, we describe the architecture of the IRSDA framework, as shown in Figure \ref{fig:irsda-system-architecture}. The architecture employs an \textit{`n-tier'} design \cite{alseelawi2020design} that incorporates \textit{`client-server (CS)'} \cite{nyabuto2023architectural} and \textit{`multi-agent system (MAS)'} \cite{merabet2014applications} models. This approach enables modularity and supports seamless integration of independent software and hardware components at each tier. We chose the CS model as it aims to efficiently distribute service requests and resources across computing devices running different IRSDA components. This design helps ensure each component meets its Service-Level Objectives (SLO) and contributes to the overall Service-Level Agreement (SLA) of the IRSDA framework. The MAS model further enhances the system by allowing distributed processing, where each agent manages separate tasks towards a unified goal of maintaining effectiveness of the whole enterprise environment.


\begin{figure*}[!htbp]
    \centering
    \includegraphics[width=\textwidth,height=0.86\textheight]{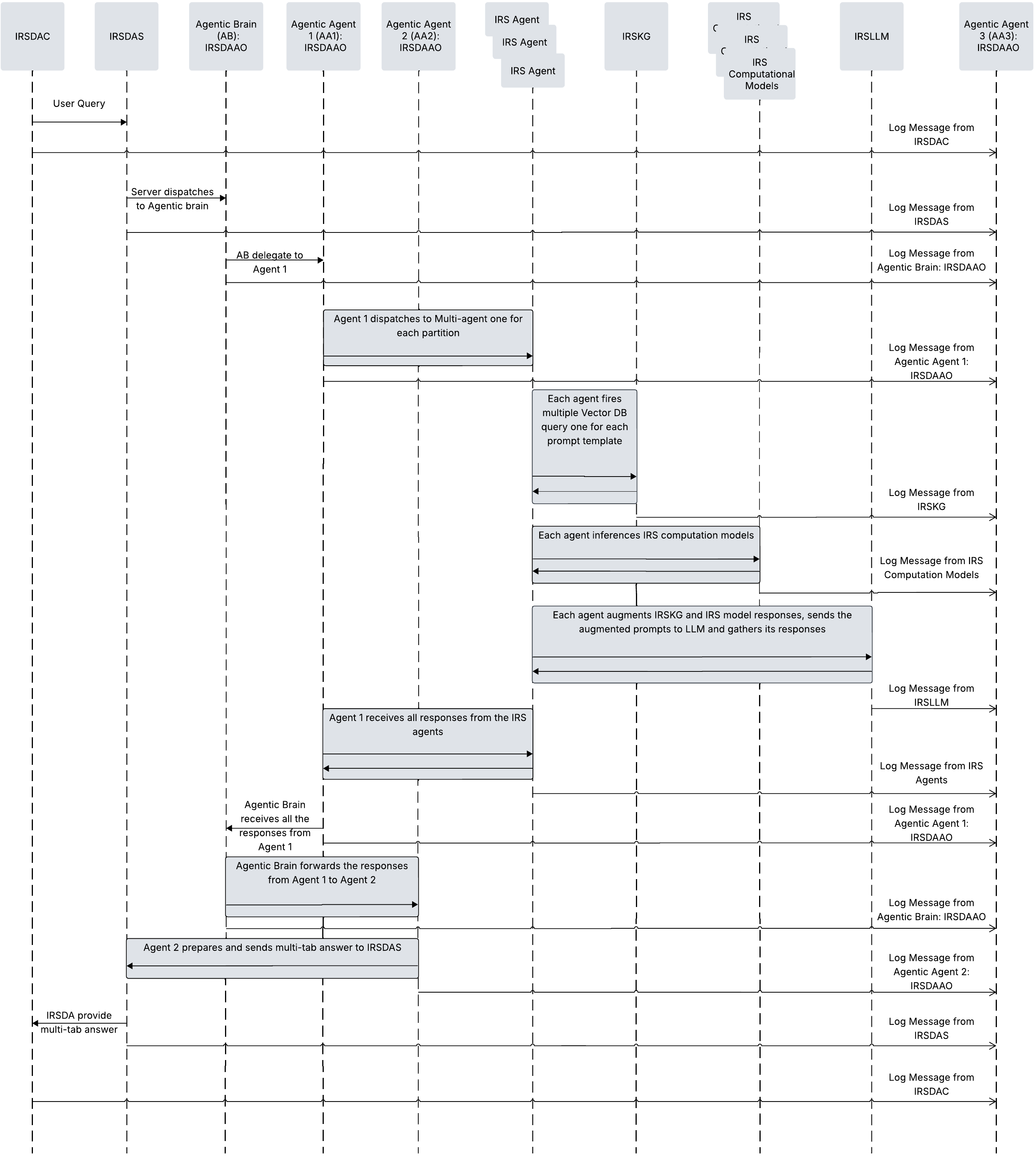}
    \caption{IRSDA Question-Answer Control Flow. The user initiates a query via the IRSDA Client (\textit{IRSDAC}), which is routed to the IRSDA Server (\textit{IRSDAS}). The query $Q_i$ is dispatched to the Incident Response System Digital Assistant Agentic Orchestration layer (\textit{IRSDAAO}), where the Agentic Brain (AB) coordinates delegation to partition-specific agents ${\p_i}$. Each agent generates prompts ${\PT_i}$ using predefined templates, retrieves contextual data from the IRS Knowledge Graph (\textit{IRSKG}), and queries the IRSLLM for responses ${\A_i}$. Agents also engage IRS computational models (e.g., GAN, RL) to evaluate potential incident response actions. The aggregated outputs are returned to IRSDAAO, which forwards the structured responses to IRSDAS. IRSDAS collaborates with IRSLLM to synthesize and format a multi-tab answer set, ultimately returned to the client interface.}
    \label{fig:IRSDA-sequence-diagram}
\end{figure*}

IRSDA system consists of multiple architectural tiers. At Tier I, the web-based IRS Digital Assistant Chatbot (IRSDAC) serves as the client-facing (dedicated for security analyst as users) interface and enables users to interact with the system using natural language. Tier II hosts the IRS Digital Assistant Server (IRSDAS), which manages client sessions, maintains state, and coordinates response activities. Tier III is dedicated to the IRS Digital Assistant Agentic Orchestration layer (IRSDAAO), where the central orchestration component delegates tasks to specialized agents. In Tier IV, IRS agents manage individual system components. Tier V includes the IRS Knowledge Graph (IRSKG) \cite{irskg} and the IRS Large Language Model (IRSLLM). The IRSKG stores enterprise logs, configuration data, and security rules, while the IRSLLM is fine-tuned on both enterprise-specific~\cite{mitra2024localintel} and public cybersecurity datasets.

Figure \ref{fig:IRSDA-sequence-diagram} illustrates the system’s control flow. When a user initiates a query through IRSDAC, the request is routed to IRSDAS and then delegated to the orchestration layer. The Agentic Brain (AB) coordinates the assignment of tasks to partition-specific agents, which retrieve contextual information from IRSKG and generate responses using IRSLLM. Computational models, such as those based on graph neural networks or reinforcement learning, can also be invoked to support incident response. The orchestrated results are returned through the system, summarized, and presented to the user in a structured format.

The architecture supports both knowledge extraction into IRSKG and consumption by IRSLLM and other system modules. Knowledge extraction involves ingesting and indexing data such as logs, configurations, and security rules, while knowledge consumption refers to the retrieval of relevant information during incident response. Additional details about the IRSKG and IRSLLM components are provided in Sections \ref{Knowledg-Graph} and \ref{llm}, respectively.

Having established the overall system architecture and described the organization of its core components, we now turn to the data foundation that underpins the IRSDA system. The following section introduces the IRSKG, which serves as the central repository for enterprise information and enables advanced retrieval capabilities.

\subsection{Digital Assistant Data Source - Knowledge Graph} \label{Knowledg-Graph}
In this section, we discuss the IRS Knowledge Graph, which functions as a graph-enabled database for enterprise textual information. We model the entire enterprise system as a graph $\G$, representing complex relationships among assets and configurations. Vector indices are created to enable efficient retrieval of relevant information during user queries, particularly when using the retrieval-augmented generation (RAG) technique (Section \ref{rag}). The IRSKG stores all relevant enterprise data as a set of \textit{`vertices'} ($\V$) and \textit{`edges'} ($\E$), each with associated \textit{`properties'} and \textit{`labels'}, as formalized in Equation \ref{eq:graph-ve}.

\begin{subequations}\label{property-graph}
    \begin{equation} \label{eq:graph-ve}
             \G = \ (\V, \ \E) \ | \ \E_{i,j} \in \E,  \ \V_i \in \V  
    \end{equation} 
    \begin{equation} \label{eq:graph-v}
        \begin{array} {ll}
             \V_i = \{\lb(\V_i), \ \pr(V_i) \} \\
        \end{array}    
    \end{equation}     
    \begin{equation} \label{eq:graph-e}
        \begin{array} {ll}
             \E_{i,j} = \{ \lb(\E_{i,j}), \pr(\E_{i,j})\}) \\
        \end{array}    
    \end{equation}     
\end{subequations}

A summary of all notation used in this paper is provided in Table \ref{tbl:notation-table}. The IRSKG schema stores various types of enterprise information, including system configurations, logs, ROE, and model training input data, which are further elaborated in the upcoming subsubsections. To begin, we examine how the IRSKG represents enterprise system configurations.

\subsubsection{Enterprise Systems Configurations} \label{system-config}
The IRSKG schema encodes the configuration of the enterprise system $\M$. We model each \textit{component type} as an index $i$, which corresponds to a template or a class in the object oriented concept. Examples of $i$ include hardware devices, virtual appliances, software modules, web servers, database servers, network switches, and container images. A \textit{component} is an instance of a type whereas we define a \textit{partition} as the set of all the components of a given type $i$, i.e., $\p_{i} = \cup_{j=1}^m i_j$, where $i_j$ represents component $j$ of type $i$, and $m$ is the total number of components of type $i$. The system $\M$ is the set of all partitions, i.e., $\M=\{{\p_1, \p_2, \dots, \p_n}\}$, where $n$ is the total number of partitions. This partitioned representation enables IRSDA to employ multiple agents, each operating locally on a single partition $\p_i$ to optimize its cybersecurity defense, while the collection of agents provides coordinated protection for the entire system $\M$. With the configuration data in place, we now consider the storage and utility of system logs within the knowledge graph.

\subsubsection{Enterprise Systems Logs} \label{system-logs}
The IRSKG serves as an \textit{`enterprise data warehouse'} for $\M$, ingesting $\M$’s system logs and enabling visualization and analysis of their relationships with other data types in accordance with Equation~\ref{eq:graph-ve}. The implementation is illustrated in Section~\ref{case-studies}. Next, we address how the IRS Knowledge Graph incorporates rules that guide and constrain the system’s automated responses.

\subsubsection{Intrusion Response System Rules of Engagement (ROE)} \label{roe}
The IRSKG also stores all IRS governing rules ($\RoE$), which encode both the conditions under which IRSDA initiates incident response and the constraints that prohibit actions deemed unsafe for $\M$. Formally, as specified in Equation~\ref{rulesengine-graph}, $\RoE={\RoE_i}$ denotes the set of all governing rules. Finally, we turn to the inclusion of data used to train and inform the system’s computational response models.

\begin{subequations}\label{rulesengine-graph}
    \begin{equation} \label{eq:rule-set}
        \begin{array} {ll}
             \RoE = \{{\RoE_i \ \forall \ 0 < i < n\}} 
        \end{array}  
    \end{equation}         
    \begin{equation} \label{eq:rule-graph-ve}
        \begin{array} {ll}
             \RoE_i = \{{\V_a(\RoE_i), \E(\RoE_i), \V_b(\RoE_i)\}} 
        \end{array}    
    \end{equation} 
    \begin{equation} \label{eq:rule-graph-v}
        \begin{array} {ll}
             \V_a(\RoE_i) = \{ \lb(\V_a(\RoE_i)), \pr(\V_a(\RoE_i)) \} 
        \end{array}    
    \end{equation}     
    \begin{equation} \label{eq:rule-graph-e}
        \begin{array} {ll}
             \E(\RoE_i) = \{ \lb(\E(\RoE_i)), \ \pr(\E(\RoE_i))) 
        \end{array}    
    \end{equation}     
\end{subequations}

\subsubsection{Response Computational Model Training Input Data} \label{model-training-input}
Input data for response-oriented computational models is stored in IRSKG. These models propose actions on $\M$ upon detecting cybersecurity breaches, subject to the governing constraints in $\RoE$. The models can be trained via different ML approaches. Because $\M$ is represented as a graph $\G=(\V,\E)$ in IRSKG, GNNs~\cite{mitra2024use} are particularly suitable for this scenario. Following the transformation in Equation~\ref{eq:gnn-sensor-graph-transform}, each vertex $\V_i$ maintains a property map $\pr(\V_{i_p})$ containing the feature \texttt{count}, defined as the number of to- or from-connections incident on $\V_i$. For simplicity and downstream message passing, $\pr(\V_{i_p})$ (including \texttt{count}) is also embedded within the edge property dictionary $\pr(\E_{i,j_p})$.

\begin{subequations}\label{eq:gnn-sensor-graph-transform}
    \begin{equation} \label{eq:gnn-sensor-graph-transform-vertex}
        \begin{array} {ll}
             \pr_{count}(\V_i)  = deg(\V_i) \ | \ deg \textrm{\ = degree of vertex $\V_i$}\\
        \end{array}  
    \end{equation}
    \begin{equation} \label{eq:gnn-sensor-graph-transform-edge}
        \begin{array} {ll}
             \pr_{count}(\E_{i,j}) = \pr_{count}(\V_i) + \pr_{count}(\V_j)
        \end{array}    
    \end{equation}     
\end{subequations}

Having established the structure, content, and capabilities of the IRSKG, we now proceed to examine the generative AI model component. The following subsection describes the design and specialization of the LLM that utilizes this data foundation for advanced security reasoning and automated response.

\subsection{Intrusion Response System (IRS) Generative AI Model} \label{llm}
    In our system, LLMs are designated to automate incident triage, prioritize response actions, and support root cause investigations by synthesizing logs, network traces, and related telemetry stored in the IRSKG enterprise data warehouse. They also provide a natural language interface that enables analysts and operations staff to query and act on security context efficiently. Several security specialized LLMs exist, such as Garak~\cite{derczynski2024garak}, CySecBERT~\cite{bayer2022cysecbert}, and SecureBERT \cite{llmsecuritytool, llmsecmodels}. We consider CyBERTuned \cite{jang-etal-2024-ignore} training approach, as it is trained using CyBERT~\cite{ranade2021cybert}, a cybersecurity text classifier, and thus directly benefits from pretraining for improved domain-specific understanding. We further adapt this model to the enterprise by fine tuning it on curated IRSKG corpora using low rank adaptation (LoRA) \cite{hu2021lora}. The resulting model is the Intrusion Response System Security Large Language Model, or IRSLLM. IRSLLM is grounded in high quality enterprise knowledge to reduce hallucinations and to improve factual fidelity in line with guidance on LLM grounding \cite{kenthapadi2024grounding}. We also mitigate hallucination risk through explicit task design and prompt templates that reflect operational intent, as described in Section~\ref{llm-tasks}. Prompt templates are whitelisted~\cite{pawelek2025llmz+} prior to selection to ensure security against injection attacks. In addition, we employ RAG over IRSKG to ground outputs with factual sources, as described in Section~\ref{rag}. Having introduced the IRSLLM, we next examine its core functional tasks within the IRSDA framework in the next subsection.

\subsection{Intrusion Response System Security Large Language Model (IRSLLM) Tasks} \label{llm-tasks}
    IRSLLM is capable of standard LLM tasks such as summarization, information retrieval, and conversational interaction \cite{radford2019language, naveed2023comprehensive}. Within the IRSDA framework, however, it is restricted to two primary tasks: question answering and summarization and can be enforced via prompt whitelisting \cite{pawelek2025llmz+}. IRSDAC (client) employs the former to interact with enterprise security personnel, whereas IRSDAS (server) uses the latter to synthesize comprehensive responses to user submissions. 
    
    To guide model output, we maintain a set of prompt templates ${\PT}$, with one template per prompting strategy $\PT_i$ (multi-shot, zero-shot, and chain-of-thought) \cite{promptEng}. For a given user query $\Q_i$, IRSDAS instantiates the appropriate $\PT_i$ with contextual evidence via RAG (Section~\ref{rag}) to improve relevance and reduce hallucinations \cite{izacard2022few}. IRSLLM then returns a preliminary answer $\A_i$ for each instantiated template; these answers are subsequently summarized $\U_i$ by IRSLLM and ensembled into the candidate set ${\A}$. As depicted in Figure~\ref{fig:IRSDA-sequence-diagram}, the resulting candidates are delivered to the user as alternative, tabbed responses. To support these tasks with reliable, contextually relevant information, the following subsection details our graph-based RAG pipeline.

\subsection{Graph Retrieval Augmented Generation (GRAG)} \label{rag}
    IRSDA adopts graph retrieval-augmented generation (GRAG) \cite{hu2024grag, edge2024local}, a form of retrieval-augmented generation (RAG) \cite{lewis2020retrieval, rag-2, promptEng}, to supply IRSLLM with grounded evidence from the IRSKG data warehouse and thereby reduce hallucination. For a user query $\Q_i$, the system retrieves relevant data chunks from IRSKG, instantiates the appropriate prompt template $\PT_i$ with those chunks, and submits the augmented prompt to IRSLLM to obtain a response $\A_i$. For summarization, IRSDA instantiates $\PT_i$ with the preliminary answer $\A_i$ and supporting chunks newly retrieved from IRSKG, and IRSLLM produces the summary $\U_i$. For both tasks, the IRSKG data-consumption path is depicted in Figure~\ref{fig:irsda-system-architecture}; the data-extraction path (a prerequisite to consumption) is discussed subsequently. To enable efficient semantic retrieval of infrastructure data and policy~\cite{mitra2025falcon}, IRSKG maintains vector-based indices created during data ingestion; these indices are persisted within the knowledge graph and used to return relevant chunks via vector database search \cite{jing2024large}. With the RAG pipeline in place, we now turn to the orchestration of agentic AI workflows that coordinate these components across the IRSDA system.

\subsection{Incident Response System Digital Assistant Agentic Orchestration (IRSDAAO)} \label{agentic}
    In this subsection, we present an overview of our IRSDAAO implementation, which leverages the Google Agent Development Kit (ADK) \cite{google-adk} for agent orchestration. While our approach is platform-agnostic, we selected ADK due to its alignment with the practical requirements of our system. The native support for audio and video modalities allowed us to expose multi-modal capabilities to end-users through the IRSDAC interface. Additionally, ADK streamlined our development process by reducing library dependency issues and offering a sandbox environment that allowed us to orchestrate agentic behaviors without diverting focus to lower-level engineering tasks. Hence, we chose ADK to prototype our agentic workflows in a controlled and reproducible setting.

    Our IRSDAAO architecture adopts a multi-agent system (MAS) design centered around an orchestration component we refer to as the \textit{`agentic brain' (AB)}. Our \textit{AB} coordinates multiple specialized subagents and manages both task routing and control flow. In our current implementation, the system comprises three subagents (as shown in Figure \ref{fig:IRSDA-sequence-diagram}): one responsible for executing user requests, a second for synthesizing results, and a third for monitoring and logging IRSDA system functionality. Our architecture is designed to be extensible, allowing for the integration of additional agents as needed. In the next section, we explore the implementation of IRSDA in a practical case study \ref{case-studies}.

\section{Case Study Intrusion Response System Digital Assistant (IRSDA) System - Online Boutique}  \label{case-studies}
    To demonstrate the practical applicability of the IRSDA architecture described in Section~\ref{irsda-design}, we delineate its components within the context of an e-commerce enterprise system. Specifically, we leverage the open-source Online Boutique (OB) 2.0 web application by Google, which serves as a representative microservice-based online retail platform comprising eleven distinct services \cite{ob}. This case study uses the AICA prototype by Blakely et al. \cite{aica-agent}, Neo4j~\cite{neo4j} as IRS Knowledge Graph (IRSKG), and Graylog~\cite{graylog} for collecting router logs.
    
    We structure this case study around a hypothetical security breach scenario in which the \texttt{front-end service} of OB is compromised, enabling an attacker to send unsolicited emails by abusing the \texttt{email service}. The IRSDA system is deployed to fulfill two core objectives: first, to prevent any further malicious email transmission, and second, to restore the \texttt{frontend service} to its intended secure state. For clarity of exposition, we focus on the containment and prevention task, assuming a ROE policy that explicitly denies network communication between the compromised and target services. We walk through the process an enterprise analyst would follow, using IRSDA to investigate the breach and IRSDA’s response mechanisms. In the following subsections, we present the OB knowledge graph representation, demonstrate embedding and retrieval workflows, illustrate IRSLLM-driven response generation, discuss the fine-tuning of the LLM, and detail the orchestration workflow within IRSDAAO.

\subsection{Online Boutique Knowledge Graph} \label{OB-Knowledg-Graph}
    We begin by highlighting the OB knowledge graph, following the data modeling approach described in Section \ref{Knowledg-Graph}. We illustrate IRSKG representation of OB \texttt{front-end-\newline partition}, $\p_i$, representation out of many system partitions as shown on Github \cite{ob}. List \ref{list:ob-system-config} (in JSON format) describes the \texttt{frontend-partition} $\p_1$, of a component type, \texttt{image}, $i=1$ with components $\p_{1_{1}}$, and $\p_{1_{2}}$. $\p_1$ is a \texttt{container} type with requirements described in $f_1$.
    
\begin{figure}[ht]
    \centering
    \includegraphics[scale=0.543]{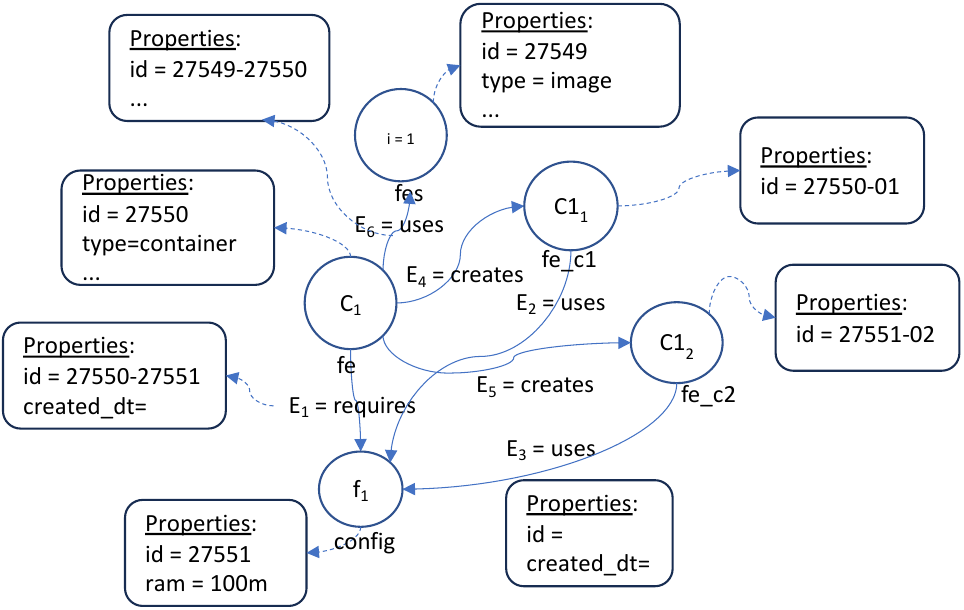}
    \caption{IRSKG representation of OB system: \texttt{frontend-partition} $\p_1$ (\texttt{containers}) of component type $i=1$ (container \texttt{images}) with configuration $f_1$ having components $\p_{1_{1}}$ and $\p_{1_{2}}$}
    \label{fig:ob-system-config}
\end{figure}

    Figure \ref{fig:ob-system-config} represents IRSKG of the scenario; we describe a part of the figure for brevity. The KG has four vertexes $\V=\{\V_1, ..., \V_4\}$ and six edges $\E=\{\E_1, ..., \E_6\}$, where the vertices are: $\V_1=\p_1$, $\V_2=f_1$, etc. Each vertex, $\V_i$, has properties such as $\p_1$ has $\pr(\V_1)=\{\textit{id=27550}\}$ with label $\lb_1=fe$. The $\p_1$ requirements of $f_1$ are represented by edge $\E_1$ with properties $\pr(\E_{i,j})=\{\textit{id=27550-27551},created\_dt=\textit{2024-10-01}\}$. $f_1$ has label $\lb_2=config$ with properties $\pr(\V_2)=\{ram=100m, id=...\}$.  

\begin{lstlisting} [language=YARA, mathescape=true,
    caption={Sample OB system configuration for the \texttt{frontend-partition} ($i=1$, \texttt{image}) showing components $p_{1_{1}}$ and $p_{1_{2}}$. The corresponding IRSKG representation appears in Fig.~\ref{fig:ob-system-config}.}, 
    label=list:ob-system-config]
{"frontend-services": 
{"type": "image"},
"partitions": {"frontend-partition": {"component-type": "container", "service_name": "frontend-service", "configuration": {"ram": "200m"}, "components":["$C1_1$","$C1_2$"]}}}
\end{lstlisting}

\begin{lstlisting} [language=YARA, mathescape=true, 
    caption={Neo4j IRSKG partition config for the JSON in Listing~\ref{list:ob-system-config}; see Fig.~\ref{fig:ob-system-config} for the graph. Nodes \textit{id=27500} and \textit{id=27549} represent \texttt{frontend-partition} ($\p_1$, \texttt{image}, $i=1$), linked by a \texttt{uses} relation with time properties.}, 
    label=list:ob-partition]
{"type":"node","id":"27500","labels":["$C_1$"],"properties":{"type":"container"}}
...
{"type":"node","id":"27549","labels":["1"],"properties":{"type":"image"}}
...
{"type":"relationship","id":"275..","label":"uses","start":{"id":"27500","properties":{"time":"23:10:45","time_month":10,"time_year":2023,"time_date":25},"end":{"id":"27549","properties":...}}
\end{lstlisting}

    Next, we illustrate the Neo4j IRSKG representation of OB system logs as shown in Listing \ref{ob-graylog-neo4j}. We demonstrate a  network interaction between a \texttt{frontend-partition} component, $\p_{1_{1}}$, and a \texttt{recommendation-partition} component $\p_{2_{1}}$. These components \texttt{runs} on hosts represented as vertexes $IP_1$ and $IP_2$ with IP \texttt{192.168.1.100} and \texttt{192.168.1.101} respectively. We only explain the first round trip communication for brevity. $IP_1$ initiates the communication by sending a \texttt{SYN}. $IP_2$ responds with a \texttt{SYN-ACK}. 

\begin{lstlisting} [language=YARA, mathescape=true, 
    caption={Sample GrayLog System Logs: TCP Handshake between Service Containers. Raw log entries capturing a TCP handshake between \textit{192.168.0.100} (hosting $\p_{1_{1}}$: \textit{frontend-service}) and \textit{192.168.0.101} (hosting $\p_{2_{1}}$: \textit{recommendation-service}).}, label=list:ob-graylog]
[2023-10-25 11:10:45] 192.168.1.100 -> 192.168.1.101: TCP SYN 
[2023-10-25 11:10:46] 192.168.1.101 -> 192.168.1.100: TCP SYN-ACK 
[2023-10-25 11:10:47] 192.168.1.100 -> 192.168.1.101: TCP ACK
[2023-10-25 11:10:48] 192.168.1.101 -> 192.168.1.100: TCP ACK
\end{lstlisting}

\begin{lstlisting} [language=YARA, mathescape=true, 
    caption={Neo4j IRSKG Graylog system mapping for the raw entries in Listing~\ref{list:ob-graylog}. Nodes $id=0$ and $id=1$ represent source (\texttt{192.168.1.100}) and destination (\texttt{192.168.1.101}), linked by a \texttt{SYN} relation with time-stamped properties.}, 
    label=ob-graylog-neo4j]
{"type":"node","id":"0","labels":["IP1"],"properties":{"ip":"192.168.1.100"}}
...
{"type":"node","id":"1","labels":["IP2"],"properties":{"ip":"192.168.1.101"}}
...
{"type":"relationship","id":"0","label":"SYN","start":{"id":"0","properties":{"time":"23:10:45,"time_month":10,"time_year":2023,"time_date":25},"end":{"id":"1","properties":...}}
\end{lstlisting}

\begin{figure}[ht]
    \includegraphics[scale=0.5]{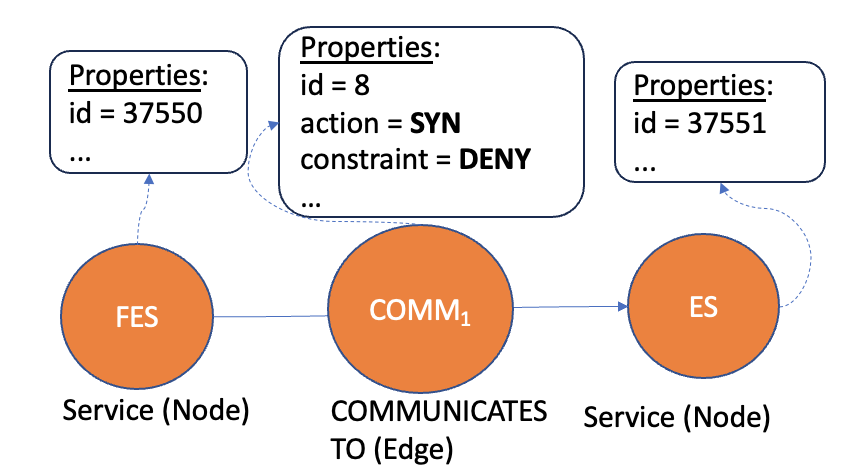}
    \caption{IRSKG representation of an OB ROE ($\RoE_1$) prohibiting \texttt{frontend-service} to \texttt{backend-service}}
    \label{fig:ob-roe}
\end{figure}

    Next, we illustrate an enterprise admin configured rule, $\RoE_1$, that prohibits \texttt{frontend-service} to communicate to \texttt{email-service}. We show in Figure \ref{fig:ob-roe} and Listing \ref{list:ob-roe}, $\RoE_1$, \textit{`deny'} any \textit{SYN}, represented as an edge type \texttt{COMMUNICATES\_TO} from \texttt{FES} to \texttt{ES} as vertices type \texttt{Service}.

    Furthermore, we demonstrate the input data to train a GNN, our computational model of choice. We illustrate how the input Gray logs, as shown in the Listing \ref{ob-graylog-neo4j}, are transformed to the GNN input data schema as shown in the Listing \ref{list:network-case-study-ingestion-phase-datatransform-neo4j-vertex-count}. We calculate the \textit{`count'} property value, $\pr(\V_{i_{p=count}})$ as $2$, following the Equation \ref{eq:gnn-sensor-graph-transform-vertex} because there are two nodes, where the ip, \textit{`192.168.1.100'} appears as either the \textit{`source'} or \textit{`destination'} address in Listing \ref{list:ob-graylog}. The source node is represented as $IP_1$ and the destination node as $IP_2$. Similarly, we assign the same \textit{`count'} value to the ip \textit{`192.168.1.101'} as shown in the Listing \ref{list:network-case-study-ingestion-phase-datatransform-neo4j-vertex-count}. Moreover, we calculate the property value, $\pr_{i,j_{p=count}}$ as two of the edge, $\E_{1,2}$ between $IP_1$ and $IP_2$, as shown in the Listing \ref{list:network-case-study-ingestion-phase-datatransform-neo4j-edge-count} abiding to the Equation \ref{eq:gnn-sensor-graph-transform-edge}.

\begin{lstlisting} [language=YARA, mathescape=true, 
    caption={Neo4j IRSKG rule: \texttt{frontend-service} is denied \texttt{CONNECT} action to \texttt{email-service}. Nodes \textit{id=27751} (\texttt{frontend-service}) and \textit{id=27750} (\texttt{email-service}) are linked by a \texttt{COMMUNICATES\_TO} edge (\textit{id=0}) with a \texttt{deny} constraint.},
    label=list:ob-roe]
{"type":`"node","id":"37550",`"labels": ["FES"],"properties": {"service_name": "frontend-service"}}
{"type": "node","id": "37551","labels": ["ES"],"properties": {"service_name": "email-service"}}
{"type": "relationship","id": "0","label": "COMM1","start": {"id": "37550","labels": ["FES"],},"end": {"id": "37551","labels": ["ES"],},"properties": {"action":"SYN","constraint": "deny","id": "6ec4f95c-f4e3-4516-92c1-172cec275696"}}
\end{lstlisting}

\begin{lstlisting} [language=YARA, mathescape=true, 
    caption={Neo4j IRSKG model input for vertices $IP_1$ (\texttt{192.168.1.100}) and $IP_2$ (\texttt{192.168.1.101}), each appearing twice in Graylog (Listing~\ref{ob-graylog-neo4j}), yielding $\pr(\V_{i_{p=count}})=2$ in the \texttt{count} property per Equation~\ref{eq:gnn-sensor-graph-transform-vertex}.},
    label=list:network-case-study-ingestion-phase-datatransform-neo4j-vertex-count]
{"type": "node","id": "2891","labels": ["IP1"],"properties": {"ip_address": "192.168.1.100","count": 2}}
{"type": "node","id": "2892","labels": ["IP2"],"properties": {"ip_address": "192.168.1.101","count": 2}}
\end{lstlisting}

\begin{lstlisting} [language=YARA, mathescape=true, 
    caption={Neo4j IRSKG model input for edge $E_{1,2}$ between $IP_1$ (\texttt{192.168.1.100}) and $IP_2$ (\texttt{192.168.1.101}), both appearing twice in Graylog (Listing~\ref{list:ob-graylog}), yielding $\pr(\E_{1,2_{p=\mathrm{count}}})=2$ in the \texttt{count} property per Equation~\ref{eq:gnn-sensor-graph-transform-edge}.},
    label=list:network-case-study-ingestion-phase-datatransform-neo4j-edge-count]
{"type": "relationship","id": "878","label": "COMMUNICATES_TO","start": {"id": "2891","labels": ["IP1"],"properties": {"ip_address": "192.168.1.100"}},"start": {...,"properties": {"ip_address": "192.168.1.101"}},"properties": {"count": 2, "id": "6ec4f95c-f4e3-4516-92c2-172cec275696"}}
\end{lstlisting}

\subsubsection{Online Boutique Embedding Knowledge Graph} \label{ob-rag-prep}
    Embeddings are created and stored (also know as vector store) for the entire IRSKG, $\G$, consisting of vertices, $\V_i$, and edges, $\E_{i,j}$, as shown in Figure \ref{fig:irsda-system-architecture} in the \textit{`Knowledge Extraction'} phase. For example, this can be done for $\G$, using the LangChain API utilizing the OpenAI embedding. \cite{emebedding-lanchain}. Alternatively, one could use the native Neo4j to store the graph itself \cite{embedding-neo4j}. We illustrate langchain Neo4j vector configuration information, $f_1$ for the \texttt{frontend-partition}, shown in Figure \ref{fig:ob-system-config} in the List \ref{list:ob-config-emebedding}. \texttt{embedding\_node\_property} has the $\pr(\V_2)$ property names and \texttt{node\_label} has the $\lb_1$ value.

\begin{lstlisting} [language=Python, 
    caption={Sample OB system embedding of configuration, $f_1$, for the \texttt{frontend-partition}.}, 
    label=list:ob-config-emebedding]
from langchain... import Neo4jVector OpenAIEmbeddings
...
config_graph = Neo4jVector.from_existing_graph(
    embedding=OpenAIEmbeddings(),
    index_name=``config_index",
    ...
    text_node_properties=["ram", "name", ...],
    embedding_node_property="config_embedding",)
\end{lstlisting}

    As shown as \textit{`Knowledge Consumption'} in Figure \ref{fig:irsda-system-architecture}, IRSDA Agents use the IRSKG embeddings to gather information on queries from the enterprise security personnel using IRSDA clients, as an illustration for a query, $Q_1$, \textit{\textbf{`Can you retrieve the enterprise system configuration for the frontend-partition?'}}. Suppose that \texttt{IRSDA Agent 1} focuses \texttt{frontend-partition}, the agent queries the IRSKG embeddings to retrieve the configuration information pertinent to that partition using the Listing \ref{list:ob-config-emebedding-retrival}.

\begin{lstlisting} [language=Python, 
    caption={\texttt{AA1} focusing on partition \texttt{frontend-partition} queries the IRSKG embeddings information for system  configuration, $f_1$.}, 
    label=list:ob-config-emebedding-retrival]
result = config_graph.similarity_search("frontend partition configuration")
\end{lstlisting}

\subsection{Intrusion Response System Large Language Model Tasks}\label{ob-irsllm-tasks}
    We demonstrated the IRSLLM tasks that follows the sequence diagram \ref{fig:IRSDA-sequence-diagram}. An enterprise security analyst launches the IRSDAC in a web browser. The analyst uses the IRSDAC posts a query, $\Q_1$, \textit{\textbf{`is there active breach in the system?'}}. IRSDAS uses a query template, $\PT_1$, to populate the query and sends the query to the embedded database described in Section \ref{ob-rag-prep} for each system partition agent. The template follows either multi-shot, zero-shot, and chain of thoughts techniques depending on availability of data to construct the templates. IRSDAS then fills another query templates, $\PT_2$, with the returned IRSKG results from the IRS system partition agents, and posts it to the IRSLLM. IRSDA gathers the results from the agents and compose a tabular format response to the IRSDAC. In this case, it produces two results. The result in tabbed format, $\U_1$, \textit{\textbf{`1) yes, frontend-service is compromised. 2) yes, frontend-service is attempting to connect to the email-service which is prohibited by the admin'}}. IRSDAC displays the results after formatting in the web browser. The analyst then poses a second query, $\Q_2$, \textit{\textbf{` Describe the rule that prohibits the connection.'}}. The IRSDA follows similar sequence flow and returns an answer with verbatim of the $\RoE_1$ with the associated listing \ref{list:ob-roe}.


\subsection{IRSLLM Fine Tuning Techniques} \label{ob-irsllm-ft}
    IRSLLM delivers accurate, policy-aligned, and explainable responses by following the fine-tuning procedure described in Section \ref{llm-tasks}, which adapts the model to the organization's operational environment and security requirements. Through this targeted fine-tuning, IRSLLM can interpret enterprise policies encoded in the IRSKG and reliably generate decisions, such as issuing a \texttt{DENY} response to unauthorized connection attempts according to the governing ROE ($\RoE_1$). In our OB breach scenario, this alignment enables IRSDA to contain the attack and restore the system to a healthy state, while seamless collaboration between IRSLLM and the IRSDAAO orchestration layer ensures all incident responses are transparent, traceable to enterprise policies, and readily actionable by analysts.

\subsection{IRSDAAO Agentic Brain Design} \label{ob-irsdaao-internal}
    The IRSDAAO module delivers orchestration and agentic intelligence throughout the IRSDA system, enabling seamless integration with the multi-agent system (MAS) framework described in Section \ref{agentic}. This implementation employs the Python-based ADK \cite{google-adk}, in which a \textit{root\_agent} serves as the central agentic brain (AB) and coordinates three subordinate \textit{sub\_agents} (AA1, AA2, AA3) in a parent-child hierarchy to manage distributed workflows.

    The operational sequence is initiated when the \textit{root\_agent} receives a user request from IRSDAS. This agent delegates the task to AA1, which is tasked with coordinating IRSAgent activities and retrieving information from partitioned knowledge sources within IRSKG. After collecting the relevant data, AA1 transfers these results to the AB, which then activates AA2. The second agent, AA2, synthesizes the information and communicates with IRSLLM to generate responses. These outputs are structured in an HTML format with multiple tabs, enhancing user navigation and clarity of presentation. The system maintains context and state across interactions using the \textit{session.state} mechanism, while AA3 runs asynchronously to log telemetry and operational metadata, ensuring transparency and traceability throughout the IRSDA lifecycle.

    Through this orchestrated workflow, IRSDA is able to detect the breach in the \texttt{frontend-service}, enforce the relevant ROE, and generate timely, context-aware responses for containment. The integration of IRSKG for knowledge representation, IRSLLM for incident reasoning, and coordinated multi-agent execution ensures that the attack remains isolated and the system is promptly restored to a secure and healthy operational state. Even with modular design and compartmentalized communication, machine learning frameworks may still be compromised by adversaries~\cite{patel2025towards}. To remain within the research scope, we do not address these adversarial scenarios. Therefore, it is crucial to ensure each component is secure before deployment in real-world applications. The IRSDA platform thus demonstrates its ability to deliver adaptive, explainable, and resilient defense across the enterprise environment. The next section presents the conclusion and outlines promising directions for future research.

\begin{lstlisting} [language=Python, 
caption={Here is the sample of the general snippet code of how to design the IRSDAAO using Google ADK notations.},
label=IRSDAAO:code-snippet]
from google.adk.agents import Agent
...
APP_NAME = "IRSDAAO"
SESSION_ID = "session_code_exec_async"
...
GEMINI_MODEL = "gemini-2.0-flash"
AA1_agent = Agent(...
    instruction="""You\'re a specialist in delegating task to IRSAgents ... and receive and return results back to AB.""",
    code_executor=BuiltInCodeExecutor())
...
root_agent = Agent(...
    tools=[agent_tool.AgentTool(agent=AA1), agent_tool.AgentTool(agent=AA2), agent_tool.AgentTool(agent=AA3)])
session_service=InMemorySessionService()
session=asyncio.run(session_service.create_session(app_name=APP_NAME, user_id=USER_ID, session_id=SESSION_ID))
logging_agent = Runner(agent=AA3, app_name=APP_NAME, session_service=session_service)
...
call_agent("is there active breach in the system?")
\end{lstlisting}

\section{Conclusion and Future Work} \label{conclusion}
In this paper, we presented the Intrusion Response System Digital Assistant (IRSDA), a unified and modular framework that combines intrusion detection and automated response through the integration of agentic orchestration, partition-scoped IRS agents, a property-graph knowledge base, and a cybersecurity-tuned LLM. The system supports real-time, ROE-compliant decision-making by coordinating agents across system partitions and grounding their actions using enterprise knowledge and telemetry through RAG. IRSDA leverages the MAPE-K loop to deliver end-to-end cyber defense, from monitoring and detection to planning and execution. Our case study on a microservices-based system demonstrated that IRSDA can effectively retrieve context-specific evidence, generate explainable and policy-aligned recommendations, and contain threats without manual intervention. This highlights the value of combining LLM reasoning with agent-based control in security-critical environments.

As a next step, we propose extending IRSDA by incorporating reinforcement learning into IRS agents to support adaptive response policies over time in response to evolving attack patterns. This could improve system agility while preserving safety constraints defined by organizational policies. By unifying detection, response, and explainability, IRSDA addresses key limitations in traditional IDS and IR workflows. As cyber threats grow in complexity and speed, frameworks like IRSDA represent a necessary evolution in automated, trustworthy, and scalable enterprise defense.

\section{Acknowledgment}
This work was supported by the Predictive Analytics and Technology Integration (PATENT) Laboratory at the Department of Computer Science and Engineering, The University of Alabama at Tuscaloosa.

\bibliographystyle{ieeetr}
\bibliography{aica-irs-digital-assistant}

\end{document}